\documentclass[twocolumn]{svjour3}          
\smartqed  
\usepackage{subfigure}
\usepackage{multicol}
\usepackage{graphicx}
\usepackage{float}
\usepackage[numbers]{natbib}
\usepackage{amssymb}
\usepackage{color}
\usepackage{natbib}
\usepackage{amsmath}
\usepackage{ulem}
\usepackage{multirow}
\usepackage{geometry}

\usepackage{geometry}  
\begin{document}

\title{Higher dimensional integrable deformations of the modified KdV equation}


\author{Xiazhi Hao \and S. Y. Lou$^*$
}


\institute{Xiazhi Hao \at College of Science, Zhejiang University of Technology, Hangzhou, China\\ 
	\and S. Y. Lou$^*$ \at
	School of Physical Science and Technology, Ningbo University, Ningbo, 315211, China\\
	$^*$Corresponding author
	\email{lousenyue@nbu.edu.cn}
              }

\date{Received: date / Accepted: date}

\maketitle

\begin{abstract}
The derivation of nonlinear integrable evolution partial differential equations in higher dimension has always been the holy grail in the field of integrability.
The well-known modified KdV  equation is a prototypical example of integrable evolution equation in one spatial dimension. 
Do there exist integrable analogs of modified KdV equation in higher spatial dimensions? In what follows, we present a positive answer to this question.
In particular, rewriting the (1+1)-dimensional integrable modified KdV equation in conservation forms and adding deformation mappings during the process allow one to construct higher dimensional integrable equations. Further, we illustrate this idea with examples from the modified KdV hierarchy, also present the Lax pairs of these higher dimensional integrable evolution equations.
\end{abstract}

\section{Introduction}
Integrable systems have a significant influence on both theory and phenomenology. Important physical applications range from fluid mechanics and nonlinear optics to plasma physics and quantum gravity \cite{sbib11,abib25,mbib11}. 
The modern history of integrable systems begins with the solution of the initial value problem of the Korteweg de Vries (KdV) equation by the inverse scattering method \cite{cbib11} and with the formulation of integrable equations as the compatibility condition of two linear eigenvalue equations called a Lax pair \cite{pbib4}.
As the inverse scattering method develops, the approaches to the problem for constructing nonlinear partial differential equations to which this method applies get themselves improved \cite{bbib2,abib14}. Such nonlinear partial differential equations may be referred to, somewhat conventionally, as integrable \cite{abib26,sbib44,jbib13,cbib7,fbib11}. Of particular interest is the higher dimensional integrable equations, which contain derivatives with respect to more than two independent variables \cite{vbib5,bbib10,cbib10}.

Of course, integrable equations are not easy to find. Recently, in the paper \cite{lou}, we proposed a beautiful idea to construct Lax and symmetry integrable equations by using deformation algorithm. 
If the original equation has conservation laws, then the deformation algorithm induces a higher dimensional integrable equation for the original equation. This is a highly efficient way of generating higher dimensional integrable equations.
In this paper, we put this construction into a concrete example in order to better understand which is the essential idea that makes the construction possible.
In addition to this, we construct a full family of higher dimensional integrable hierarchy.

Our main objective is to present a detailed account of results obtained by means of the deformation for (1+1)-dimensional integrable systems.
Our purpose is not merely to give a formulation of a problem and final result, but also to explain the method in sufficient detail if it is nontrivial.
In the following section, we shall review briefly a necessary deformation algorithm  and give a complete answer to a natural question: what higher dimensional integrable equations are the deformations of the (1+1)-dimensional ones?
An concrete example is investigated in detail in section \ref{secmkdvhd}.
In sections \ref{seclax} and \ref{sechierarchy}, we discuss two topics concerning integrability associated with higher dimensional modified KdV-HD (Harry-Dym) equation, in particular, the Lax integrability and symmetry integrability.
Section \ref{secsolution} is devoted to a particular kink-like solution formulated in terms of implicitly expression.
In last section, we make some concluding remarks and mention some difficult problems.


\section{Preliminaries}\label{pre}

In this section, we briefly review the deformation algorithm, which closely follows \cite{lou} and provides a method to deform lower dimensional integrable equations to higher dimensional integrable ones on the situation in which the lower dimensional integrable equations possess the conservation laws. 
All higher dimensional integrable equations considered in the following deformation algorithm are local. This method is based on the use of conservation laws.

\bf Deformation algorithm. \it If a general (1+1)-dimen\\sional integrable local evolution equation
\begin{equation}\begin{split}
&u_t=F(u,\ u_x,\ \ldots,\ u_{xn}),\\
& u_{xn}=\partial_x^nu, u=(u_1,\ u_2,\ \ldots,\ u_m), \label{Gen}
\end{split}\end{equation}
has some conservation laws
\begin{equation}\begin{split}
&\rho_{it}=J_{ix}, \ i=1,\ 2,\ \ldots, D-1,\ \rho_i=\rho_i(u),\\
& J_i=J_i(u,\ u_x,\ \ldots,\ u_{xN}), \label{CLaw}
\end{split}\end{equation}
where the conserved densities $\rho_i$ are dependent only on the field $u$,
then the deformed $(D+1)$-dimensional equation
\begin{equation}
\hat{T}u=F(u,\ \hat{L}u,\ \ldots,\ \hat{L}^nu)\ \label{D+1}
\end{equation}
may be integrable suppose that the deformed operators depend on $D$  variables in such a way that
\begin{equation}
\hat{L}\equiv \partial_x+\sum_{i=1}^{D-1}a_i\rho_i\partial_{x_i},\
\hat{T}\equiv \partial_t+\sum_{i=1}^{D-1}a_i\bar{J}_i\partial_{x_i} \label{LT}
\end{equation}
and the deformed flows are
\begin{equation}
\bar{J_i}=J_i|_{u_{xj}\rightarrow \hat{L}^ju,\ j=1,\ 2,\ \ldots,\ N}. \label{Jbar}
\end{equation}
\rm

This algorithm leads directly to the possibility of transforming lower dimensional integrable equations to higher dimensional ones by means of conservation laws. Furthermore, the same deformation map is enjoyed by their Lax pairs.  Indeed by extending the deformation map to the Lax pairs, we give more credence to this possibility.
Suppose that Lax pair of the (1+1)-dimensional integrable local evolution equation is
\begin{eqnarray}\label{lax}
M\psi=0, N\psi=0.
\end{eqnarray}
The compatibility condition $[M,N]=MN-NM=0$ holds provided that $u$ satisfies the original equation.
Finding the Lax pair of the $(D+1)$-dimensional equation is quite straightforward which can be achieved by using the deformation relations
\begin{eqnarray}\label{rela}
\partial_x\rightarrow \hat{L},\ \partial_t\rightarrow \hat{T}.
\end{eqnarray}
If we apply \eqref{rela} to \eqref{lax}, we readily deduce the Lax pair 
\begin{eqnarray}\label{hlax}
\hat{M}\psi=0,\ \hat{T}\psi=0.
\end{eqnarray}
Strictly speaking, equation \eqref{D+1} can be derived as compatibility condition of a Lax representation in terms of the \eqref{hlax} which can be obtained from \eqref{lax} with the use of two deformation relations \eqref{rela}.


\section{Higher dimensional modified KdV-HD equations}\label{secmkdvhd}

Our main point is to construct higher dimensional integrable equations whose Lax pairs and symmetries can be found using deformation algorithm as well.
We first examine the procedure for constructing such equation on a classical example, the modified KdV equation
\begin{eqnarray}\label{mkdv}
u_t=u_{xxx}- 6u^2u_x,
\end{eqnarray}
which occurs, for instance, in the analysis of quasi-one-dimensional solids \cite{nbib1}, liquid-crystal hydrodynamics \cite{vbib6},  acoustic wave propagation in anharmonic lattices \cite{mbib19}, and in the theory of nonlinear Alfv\'en waves in a collisionless plasma \cite{tbib2}.
The modified KdV equation is characterized by the existence of an infinite number of conservation laws \cite{rbib23,mbib15} and the first three conservation laws are 
\begin{eqnarray*}
	&&(u)_t=(u_{xx}-2 u^3)_x,\\
	&&(u^2)_t=(2uu_{xx}-u_x^2-3u^4)_x,\\
	&&(u^4+u_x^2)_t=(2u_xu_{xxx}-u_{xx}^2+4u^3u_{xx}-12u^2u_x^2-4u^6)_x.
\end{eqnarray*}
Only the first two  which correspond to conservations of momentum and energy, respectively, will be of concern to us.
By a (3+1)-dimensional modified KdV-HD equation formed from deformation algorithm with two conservation laws, we mean an equation of the form
\begin{eqnarray}
\hat{T}u=\hat{L}(\hat{L^2}u-2 u^3)
\end{eqnarray}
with deformed operators 
\begin{eqnarray*}
	&&\hat{L}=\partial_x+a_1u\partial_y+a_2u^2\partial_z,\\
	&&\hat{T}=\partial_t+a_1\bar{J_1}\partial_y+a_2\bar{J_2}\partial_z
\end{eqnarray*}
and deformed flows 
\begin{eqnarray*}
	\bar{J_1}=\hat{L^2}u-2 u^3, \bar{J_2}=2u\hat{L^2}u-(\hat{L}u)^2-3 u^4,
\end{eqnarray*}
or, in more detail
\begin{eqnarray}\begin{split}\label{mkdv-hd}
u_t&= 3bu_z(\hat{L}u)^2+(u_{xx}-2 u^3+\frac32a^2u^2u_{yy}+3auu_{xy}\\&+3bu^2u_{xz}+\frac32b^2u^4u_{zz}+3abu^3u_{zy})_x+a(a^2u^3u_{yy}\\&-u^4+\frac32au^2u_{xy}+\frac94abu^4u_{zy}+\frac65b^2u^5u_{zz}\\&+2bu^3u_{xz})_y+b(b^2u^6u_{zz}-\frac35 u^5+\frac32bu^4u_{xz}\\&+\frac34a^2u^4u_{yy}+\frac95abu^5u_{zy}+au^3u_{xy})_z.
\end{split}\end{eqnarray}
An interesting remark is that the equation \eqref{mkdv-hd} should be integrable, due to its direct map via a deformation relation to integrable modified KdV equation \eqref{mkdv}. 
Nevertheless, integrability tests such as those based on the Painlev\'e property do not work for equation \eqref{mkdv-hd} because the reciprocal transformations are included in higher dimensional equation
and a leading term cannot be even identified.

The usual (1+1)-dimensional modified KdV equation \eqref{mkdv} is a simple reduction of equation \eqref{mkdv-hd} with $u_y=u_z=0$.
In analogy, we can take $u$ under consideration to depend only on $y$, then equation \eqref{mkdv-hd} is transferred into 
\begin{eqnarray}\label{uty}
u_t= a(a^2u^3u_{yy}-u^4)_y.
\end{eqnarray}
Equation \eqref{uty} can be written in conservation forms as 
\begin{eqnarray}\begin{split}\label{utyc}
&\rho_t^{\pm}=J_y^{\pm},~ \rho^{\pm}=u^{\pm 1},~ J^+=a(a^2u^3u_{yy}-u^4), \\& J^-=a[2 u^2-\displaystyle\frac{a^2}{2}(u^2)_{yy}].
\end{split}\end{eqnarray}
If $u$ merely contains one spatial variable $z$, then equation \eqref{mkdv-hd} becomes a new reciprocal link of the modified KdV equation \eqref{mkdv}
\begin{eqnarray}\label{utz}
u_t&= &3b^3u^4u_z^3+b(b^2u^6u_{zz}-\frac35 u^5)_z.
\end{eqnarray}
The first two conservation laws associated with equation \eqref{utz} are expressed in the forms
\begin{eqnarray}
\begin{split}\label{utzc}
&(u^{-1})_t=(b u^3-\frac{b^3u^3(u^2)_{zz}}{2})_z,\\
&(u^{-2})_t=-(2b^3u^3u_{zz}+3bu^2(b^2u_z^2-1))_z.
\end{split}
\end{eqnarray}
Applying the deformation algorithm to equations \eqref{uty} and \eqref{utz} with their conservation laws \eqref{utyc} and \eqref{utzc}, respectively,
then, (3+1)-dimensional modified KdV-HD equation \eqref{mkdv-hd} is again obtained which we mention without going into detail.

The consideration of $u$ being independent of $x$ leads equation \eqref{mkdv-hd} to be a (2+1)-dimensional modified KdV-HD equation
\begin{eqnarray}\begin{split}
u_t&=3buu_z(au_y+buu_z)^2+a(a^2u^3u_{yy}-u^4+\frac94abu^4u_{zy}\\&+\frac65b^2u^5u_{zz})_y+b(b^2u^6u_{zz}-\frac35u^5+\frac34a^2u^4u_{yy}\\&+\frac95abu^5u_{zy})_z.
\end{split}\end{eqnarray}
Performing a further reduction $u_y=0$, another (2+1)-dimensional modified KdV-HD equation is obtained
\begin{eqnarray}\begin{split}
u_t&=3bu_z(u_x+bu^2u_z)^2+(u_{xx}-2u^3+3bu^2u_{xz}\\&+\frac32b^2u^4u_{zz})_x+b(b^2u^6u_{zz}-\frac35u^5+\frac32bu^4u_{xz})_z.
\end{split}\end{eqnarray}
The last (2+1)-dimensional reduction 
\begin{eqnarray}\begin{split}
u_t&= (u_{xx}-2u^3+\frac32a^2u^2u_{yy}+3auu_{xy})_x\\&+a(a^2u^3u_{yy}-u^4+\frac32au^2u_{xy})_y
\end{split}\end{eqnarray}
is obtained by imposing the condition $u_z=0$.

\section{The Lax pair of the (3+1)-dimensional modified KdV-HD equation}\label{seclax}

The (3+1)-dimensional modified KdV-HD equation \eqref{mkdv-hd} is a nonlinear higher dimensional integrable local evolution equation in four independent variables.
Here integrability means the existence of a Lax pair and infinitely many higher order symmetries. The investigation of the Lax pair for equation \eqref{mkdv-hd} can be carried out under an analogous deformation map.

The modified KdV equation \eqref{mkdv} has the significant property that it results from the compatibility condition of a certain auxiliary linear problem
\begin{eqnarray*}
	&&(\partial_x^2-u^2+u_x+\lambda)\psi\equiv M\psi=0,\\
	&&(\partial_t-4\partial_x^3+6u^2\partial_x-6u_x\partial_x+6uu_x-3u_{xx})\psi\equiv N\psi=0.
\end{eqnarray*}
This Lax pair takes after the same deformation map
\begin{eqnarray*}
	\partial_x\rightarrow \hat{L},\ \partial_t\rightarrow \hat{T}
\end{eqnarray*}
the form 
\begin{eqnarray*}
	&&(\hat{L}^2-u^2+(\hat{L}u)+\lambda)\psi\equiv \hat{M}\psi=0,\\
	&&(\hat{T}-4\hat{L}^3+6u^2\hat{L}-6(\hat{L}u)\hat{L}+6u(\hat{L}u)-3(\hat{L}^2u))\psi\equiv \hat{N}\psi=0,
\end{eqnarray*}
which is exactly the Lax representation of the (3+1)-dimensional modified KdV-HD equation \eqref{mkdv-hd}.

\section{The (3+1)-dimensional integrable modified KdV-HD hierarchy and its Lax pair}\label{sechierarchy}

The existence of an infinite hierarchy of commuting flows is one of the most important properties of integrability.
All classical integrable equations such as KdV \cite{pbib7}, Kadomtsev-Petviashvili \cite{sbib15}, 
Ablowitz–Kaup–Newell–Segur \cite{sbib45}, Davey-Stewartson \cite{sbib20}, etc., were discovered as equations with infinitely many symmetries. 
For the modified KdV equation, there exists a hierarchy of higher commuting flows with related symmetries and recursion operator as well.
The modified KdV hierarchy is defined by the infinite sequence of flows with respect to the times $u_{t_{2n+1}}$, with the $n^{th}$ member given by the equation \cite{bbib12,mbib18,zbib20,sbib43,wbib27,pbib12}
\begin{equation}\begin{split}\label{mKdVH}
&u_{t_{2n+1}}=K_{2n+1}=\Phi^nu_x,\ \Phi=\partial_x^2-4\partial_xu\partial_x^{-1}u, \\&\ n=0,\ 1,\ \ldots,\ \infty.
\end{split}\end{equation}
For $n=1$, equation \eqref{mKdVH} is exactly the modified KdV equation while the fifth order modified KdV equation
\begin{eqnarray}
u_{t_{5}}=(u_{xxxx}-10uu_{x}^2-10u^2u_{xx}+6u^5)_x
\end{eqnarray}
is related to \eqref{mKdVH} with $n=2$. When $n>2$, \eqref{mKdVH} gives the higher order versions of the modified KdV equations.
For the purpose of deforming the $n^{th}$ order modified KdV-HD equation, we first introduce the
conserved densities $u$ and $u^2$  with the differential polynomial flows in the forms
\begin{eqnarray}
u_{t_{2n+1}}&=&(J_{2n+1})_x,\ J_{2n+1}=\partial_x^{-1}\Phi^nu_x,\ \nonumber\\
(u^2)_{t_{2n+1}}&=&(G_{2n+1})_x,\ G_{2n+1}=2\partial_x^{-1}u\Phi^nu_x,\  \label{JGn}
\end{eqnarray}
where $J_{2n+1}$ and $G_{2n+1}$ are all differential polynomials of $u$ with respect to $x$. For $n=2$, \eqref{JGn} yields
\begin{eqnarray}\begin{split}
J_{5}&=u_{xxxx}-10uu_{x}^2-10u^2u_{xx}+6u^5, \\
G_{5}&=2uu_{xxxx}-2u_xu_{xxx}+u_{xx}^2-20u^3u_{xx}\\&-10u^2u_x^2+10u^6.\label{JG5}
\end{split}\end{eqnarray}
On the basis of the conservation laws \eqref{JGn} and the deformation algorithm,
the whole (3+1)-dimensional modified KdV-HD hierarchy follows \eqref{mKdVH} directly as $\partial_t\rightarrow \hat T_{2n+1}, u_{xm}\rightarrow \hat{L}^mu$ 
\begin{eqnarray}\label{hmkdvh}
\hat{T}_{2n+1}u=(\Phi^nu_x)|_{u_{xm}\rightarrow \hat{L}^mu}, m=1, 2, \ldots, 2n+1,  
\end{eqnarray}
where $ \hat{T}_{2n+1}=\partial_{t_{2n+1}}+a_1\bar{J}_{2n+1}\partial_y+a_2\bar{G}_{2n+1}\partial_z, ~~\hat{L}=\partial_x+a_1u\partial_y+a_2u^2\partial_z,~~ \bar{J}_{2n+1}=J_{2n+1}|_{u_{xm}\rightarrow \hat{L}^mu} $ and
$ \bar{G}_{2n+1}=G_{2n+1}|_{u_{xm}\rightarrow \hat{L}^mu}, m=1,\ 2,\ \ldots,\ 2n+1$.
The (3+1)-dimensional modified KdV-HD hierarchy \eqref{hmkdvh} is equivalent to
\begin{eqnarray}\begin{split}\label{3KdVH}
u_{t_{2n+1}}&=(1-a_1u_y\partial_x^{-1}-2a_2u_z\partial_x^{-1}u)\\&\Phi^nu_x|_{u_{xm}\rightarrow
	\hat{L}^mu,\ m=1,\ 2,\ \ldots,\ 2n+1} \equiv \bar{K}_{2n+1}
\end{split}\end{eqnarray}
with the first three $\bar{K}_{2n+1}$ being 
\begin{eqnarray*}
	\bar{K}_1&=&u_x,\\
	\bar{K}_3&=&(\hat{L}-a_1u_y)(\hat{L}^2u-2u^3)-a_2u_z[2u\hat{L}^2u-(\hat{L}u)^2-3u^4],\\
	\bar{K}_5&=&(\hat{L}-a_1u_y)[\hat{L}^4u-10u(\hat{L}u)^2-10u^2\hat{L}^2u+6u^5]\\&
	-&a_2u_z[2u\hat{L}^4u-2(\hat{L}u)(\hat{L}^3u)+(\hat{L}^2u)^2-20u^3\hat{L}^2u\\&-&10u^2(\hat{L}u)^2+10u^6].
\end{eqnarray*}
It can be confirmed by a calculation that the compatibility condition, $u_{t_{2n+1}t_{2m+1}}=u_{t_{2m+1}t_{2n+1}}$, of the (3+1)-dimensional modified KdV-HD hierarchy \eqref{3KdVH} is clearly satisfied, i.e.,
\begin{eqnarray*}
&&[\bar{K}_{2n+1},\ \bar{K}_{2m+1}]=\lim_{\epsilon=0} \frac{\mbox{d}}{\mbox{d}\epsilon}[\bar{K}_{2n+1}(u+\epsilon \bar{K}_{2m+1})\\&&- \bar{K}_{2m+1}(u+\epsilon \bar{K}_{2n+1})]=0.
\end{eqnarray*}

The family of the KdV hierarchy is generated by making use of the recursion operator 
\begin{eqnarray}
\Psi=\partial_x^2+4v+2v_x\partial_x^{-1}
\end{eqnarray}
in the differential relation
\begin{eqnarray}\label{kdvh}
v_{t_{2n+1}}=\Psi^{n}v_x,\quad n=0,1,2,\dots.
\end{eqnarray}
The KdV hierarchy is also recognized as the solvability condition for the Lax representation \cite{pbib12,xbib27}
\begin{eqnarray}\label{kdvlax1}
&&\psi_{xx}=(\lambda-v)\psi,\\\label{kdvlax2}
&&\psi_{t_{2n+1}}=\frac{\Gamma_x(v,\lambda)}{2}\psi-\Gamma(v,\lambda)\psi_x
\end{eqnarray} with
\begin{eqnarray*}
	&&\Gamma(v,\lambda)=-2\sum_{k=0}^{n}((4\lambda)^{n-k}L_k[v]), \\
	&&	\partial_x L_{k+1}[v]=(\partial_{xxx}+4v\partial_x+2v_x)L_k[v],~L_0[v]=\frac{1}{2},
\end{eqnarray*}
which holds good for all equations in the KdV hierarchy.
The nonlinear transformation of Miura or the so-called Miura transformation \cite{mbib18,abib24}
\begin{eqnarray}
v=u_x-u^2
\end{eqnarray} 
converts equations \eqref{kdvlax1}-\eqref{kdvlax2} into the Lax representation
\begin{eqnarray}\label{mkdvlax1}
&&\psi_{xx}=(\lambda-(u_x-u^2))\psi,\\\label{mkdvlax2}
&&\psi_{t_{2n+1}}=\frac{\Gamma_x((u_x-u^2),\lambda)}{2}\psi-\Gamma((u_x-u^2),\lambda)\psi_x
\end{eqnarray} with
\begin{eqnarray*}
	&&\Gamma((u_x-u^2),\lambda)=-2\sum_{k=0}^{n}((4\lambda)^{n-k}L_k[u_x-u^2]), \\
	&&	\partial_x L_{k+1}[u_x-u^2]=(\partial_{xxx}+4(u_x-u^2)\partial_x\\&&+2(u_x-u^2)_x)L_k[u_x-u^2],~L_0[u_x-u^2]=\frac{1}{2}
\end{eqnarray*}
of the whole modified KdV hierarchy \eqref{mKdVH}.   
It is of interest to note that the Lax representation of all equations in  the (3+1)-dimensional modified KdV-HD hierarchy \eqref{hmkdvh} can be identified from \eqref{mkdvlax1}-\eqref{mkdvlax2} through the deformation map $\partial_x\rightarrow \hat{L},\ \partial_t\rightarrow \hat{T}_{2n+1}.$

\section{Anomalous kink wave}\label{secsolution}

\begin{figure}
	\begin{center}
		\subfigure[]{
			\includegraphics[width=0.31\textwidth]{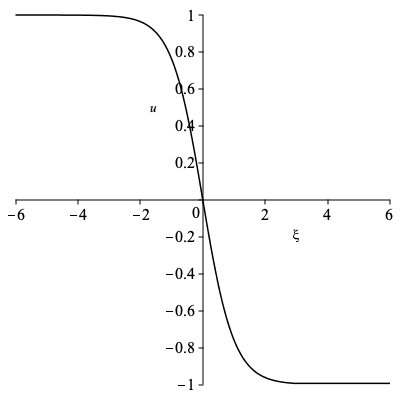}}
		\centering
		\subfigure[]{
			\includegraphics[width=0.3\textwidth]{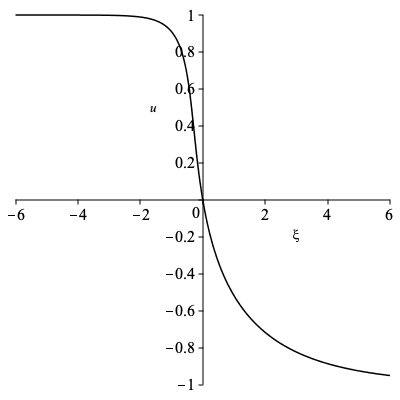}}
		\centering
		\subfigure[]{
			\includegraphics[width=0.3\textwidth]{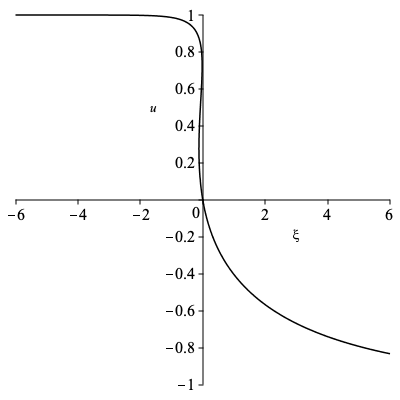}}
	\end{center}
	\caption{The images of different single kink shapes of \eqref{Soliton} under the same parameters $a=1,b=1,c=-1, k=1$ and different deformation parameters $p$ and $q$. (a). A quasi-symmetric kink with small deformation parameter selections $p=-1/100,q=1/100$. (b). An asymmetric kink with parameter selections $p=-2,q=2$. (c). An anomalous kink under the larger  parameter selections $p=-5,q=5$.}\label{figs}
\end{figure}

Despite integrability manifested in the existence of a Lax pair and higher order symmetries, the analytical explicit solutions \cite{sbib10,sbib3,rbib21,ybib17,rbib17,rbib18}, however, of the  (3+1)-dimensional modified KdV-HD equation \eqref{mkdv-hd} are not easily accessible because the direct application of the standard analytic techniques, such as B\"acklund transformation, Hirota's method, etc. \cite{abib3,abib27,abib28}, for deriving explicit solutions of the equation \eqref{mkdv-hd} fails.
Therefore, taking the simplest traveling wave solution into account, a solitary wave solution of the (3+1)-dimensional mKdV-HD equation \eqref{mkdv-hd} can be sought in the form
\begin{eqnarray}
u=U(\xi),~ \xi=kx+py+qz+\omega t+\xi_0
\end{eqnarray}
with $k,p,q,\omega, \xi_0$ constants, where $U$ satisfies a third order ordinary differential equation
\begin{eqnarray*}
\omega U_{\xi}&=&[(k+apU+bqU^2)^3U_{\xi\xi}-U^3(2k+apU\\&+&\frac35bqU^2)]_{\xi}+3bq(k+apU+bqU^2)^2U_{\xi}^3.
\end{eqnarray*}
The implicit expression 
\begin{eqnarray}\label{Soliton}
U(\xi)=-c\tanh[\frac{c(ap\ln(U-c)+bqU-\xi)}{acp-bc^2q-k}]
\end{eqnarray}
with
$\omega=bqc^4-2kc^2$
gives the single kink-like solution of equation \eqref{mkdv-hd}.
It can be seen from this expression that when $p$ and $q$ are equal to zero, leads to the standard kink solution of the usual modified KdV equation. 
As $|p|$ and $|q|$ increasing, the kink shape will deform to an anomalous kink wave. 
In Fig. \ref{figs}, the $u$ is plotted as a function of $\xi$ for various values of $p$ and $q$.
We are only able to derive an anomalous kink wave of the (3+1)-dimensional modified KdV-HD equation \eqref{mkdv-hd},  but no further multi-soliton solution.

\section{Conclusions and discussions}

In this study, we have presented a method for generating higher dimensional integrable modified KdV-HD equations by deformation algorithm which have not previously been derived.
Moreover, these higher dimensional modified KdV-HD equations turn out to be integrable in the sense that associated with them there exist compatible pairs of linear systems and infinitely many symmetries. 
The strength of the deformation algorithm is that it constructs (D+1)-dimensional integrable systems, as well as integrable hierarchies, in a straightforward manner. The whole (3+1)-dimensional modified KdV-HD hierarchy and its Lax pair have also been obtained via the deformation algorithm from the (1+1)-dimensional modified KdV hierarchy.

It is plain that the method employed in this paper provides the possibility to deform a large number of new higher dimensional integrable systems. The modified KdV equation specifically discussed above is merely an example of the vistas opened by this methodology to deform new higher dimensional integrable systems.



Finally, we give several remarks. 

First, conservation laws provide an efficient method for the construction of higher dimensional integrable systems in a regular way. 
Although we have been concerned with the modified KdV equation, this method can be applied to a wide class of nonlinear evolution equations. An analogous construction of higher dimensional integrable equations can be carried out whenever the conservation laws exist in the original integrable equations.


Second, from our viewpoint, the most important consequence of these results is that they establish a connection between lower dimensional integrable systems and higher dimensional ones. 
For any higher dimensional deformed integrable systems, there exist some new (1+1)-dimensional integrable reductions. Again, applying the deformation algorithm to these reductions, one can re-obtain the same higher dimensional deformed integrable systems.

Third, this method has the advantage that it constructs integrable higher dimensional local nonlinear systems, as well as their Lax pairs and higher order symmetries, in a straightforward manner, but it has the disadvantage that it may destroy other exact integrability.
For instance, integrability tests such as those based on the Painlev\'e property do not hold because the reciprocal transformations which account for the anomalous nature of the higher dimensional equations are included.

Last but not least, finding the exact solutions of the higher dimensional integrable equations are quite difficult even for the solitary solutions.
Of course, there are some standard analytic techniques for obtaining solutions of nonlinear integrable evolution equations, but, for the higher dimensional integrable equations, some have proved unsuccessful while others have not been attempted. Research on this issue is in progress.

\section*{Acknowledgement} 
The work was sponsored by the National Natural Science Foundations of China (Nos.  12235007, 11975131, 11435005, 12275144, 11975204), K. C. Wong Magna Fund in Ningbo University, Natural Science Foundation of Zhejiang Province No. LQ20A010009.
\\\\

\bibliographystyle{spmpsci}      
\bibliography{ref}

\end{document}